\documentclass[aps,prb,twocolumn,superscriptaddress]{revtex4-2}

\usepackage{txfonts}
\usepackage{color}
\usepackage{soul}
\usepackage{mathrsfs}
\usepackage{graphicx}

\begin{document}


\title{Uniaxial stress effects on magnetic and electric properties of the ground state in a centrosymmetric magnetic skyrmion host Gd$_2$PdSi$_3$}



\author{Hiraku Saito}
\email{h3110@issp.u-tokyo.ac.jp}
\author{Naoki Kobayashi}
\affiliation{Institute for Solid State Physics, The University of Tokyo, Kashiwanoha, Chiba 277-0882, Japan}

\author{Takuro Kawasaki}
\author{Tatsuya Nakamura}
\affiliation{J-PARC Center, Japan Atomic Energy Agency, Tokai, Naka, Ibaraki 319-1195, Japan}

\author{Akiko Kikkawa}
\author{Yasujiro Taguchi}
\author{Yoshinori Tokura}
\affiliation{RIKEN Center for Emergent Matter Science (CEMS), Wako, Saitama 351-0198, Japan}

\author{Taro Nakajima}
\affiliation{Institute for Solid State Physics, The University of Tokyo, Kashiwanoha, Chiba 277-0882, Japan}
\affiliation{RIKEN Center for Emergent Matter Science (CEMS), Wako, Saitama 351-0198, Japan}
\affiliation{Institute of Materials Structure Science, High Energy Accelerator Research Organization, Tsukuba, Ibaraki 305-0801, Japan}


\date{\today}

\begin{abstract}
We investigate effects of uniaxial stress to magnetic orders and electrical resistivity of the centrosymmetric magnetic skyrmion compound Gd$_2$PdSi$_3$, which has a hexagonal crystal structure composed of triangular lattice layers of magnetic Gd$^{3+}$ ions. %
This compound is known to exhibit the triple-$q$ magnetic skyrmion lattice phase with a giant topological Hall effect
in the first field induced phase [T. Kurumaji \textit{et al}. Science \textbf{365}, 914-918 (2019)]. %
In contrast to the established picture of the field-induced phase, the ground state of this system still remains to be studied. %
Although previous studies reported the existence of the incommensurate magnetic modulations described by a magnetic modulation wave vector ${\bf q}=(q,0,0)$ where $q\sim 0.14$ and its equivalents, it is still unclear whether the magnetic structure is a single-$q$ structure or a multiple-$q$ structure. %
In the present study, we performed magnetization, resistivity and neutron diffraction measurements with a compressive uniaxial stress applied perpendicular to the $c$ axis. %
The observed data revealed that the system did not exhibit anisotropic magnetic and electric properties expected from a single-$q$ magnetic order, suggesting that the magnetic ground state of this system is a multi-$q$ magnetic order. %
\end{abstract}


\maketitle

\section{Introduction}
Magnetic skyrmions, which are vortex-like spin arrangements having nontrivial topology, have been extensively studied in condensed matter physics since the discovery of the magnetic Skyrmion lattice (SkL) in the chiral magnet MnSi\cite{muhlbauer_2009} and Fe$_{1-x}$Co$_x$Si\cite{yu_2010}.
Whereas the SkLs had been initially studied in magnetic materials with broken spatial inversion symmetry, in which the Dzyaloshinskii-Moriya (DM) interactions favor non-collinear magnetic modulations including the SkLs, the discovery of SkL phase in the centrosymmetric intermetallic compound Gd$_2$PdSi$_3$\cite{kurumaji_2019} demonstrated that the SkLs can be realized without relying on the DM interaction. %
Thus far, a variety of rare-earth based centrosymmetric magnets, such as GdRu$_2$Si$_2$\cite{GRS_Khanh_NPhys}, GdRu$_2$Ge$_2$\cite{GdRu2Ge2_Yoshimochi_2024}, Gd$_3$Ru$_4$Al$_{12}$\cite{GRA_Max_NCom}, EuAl$_4$\cite{EuAl4_Takagi_2022} etc., have been reported to display SkL phases and related topological spin textures. %

Various theoretical models have been proposed for the stabilization of SkL phase in centrosymmetric systems; for instance, magnetic frustration \cite{PRL_Okubo_MonteCarlo,NCom_Leonov_frustratedmag}, couplings between conduction electrons and magnetic moments\cite{PRB_Hayami_biquadratic,PRL_Wang_2DEG}, anisotropic exchange interactions\cite{Sk_AnisoExchange_NJP_2021}, dipole-dipole magnetic interactions\cite{Sk_dipolar_PRB_2021} and so on. %
These models can lead to magnetic orders described by multiple magnetic modulation wavevectors ($q$-vectors) which are referred to as multi-$q$ orders. %
To pin down the microscopic mechanism for stabilizing the SkLs in the centrosymmetric compounds, it is important to experimentally determine the magnetic structures not only for the SkL phases but also in the other magnetic phases appearing with varying temperature and magnetic field. %
More specifically, it is necessary to experimentally make a distinction between multi-$q$ and single-$q$ magnetic structures, for each magnetic phase. %

\begin{figure}[t]
	\includegraphics[keepaspectratio, width=8cm]
	{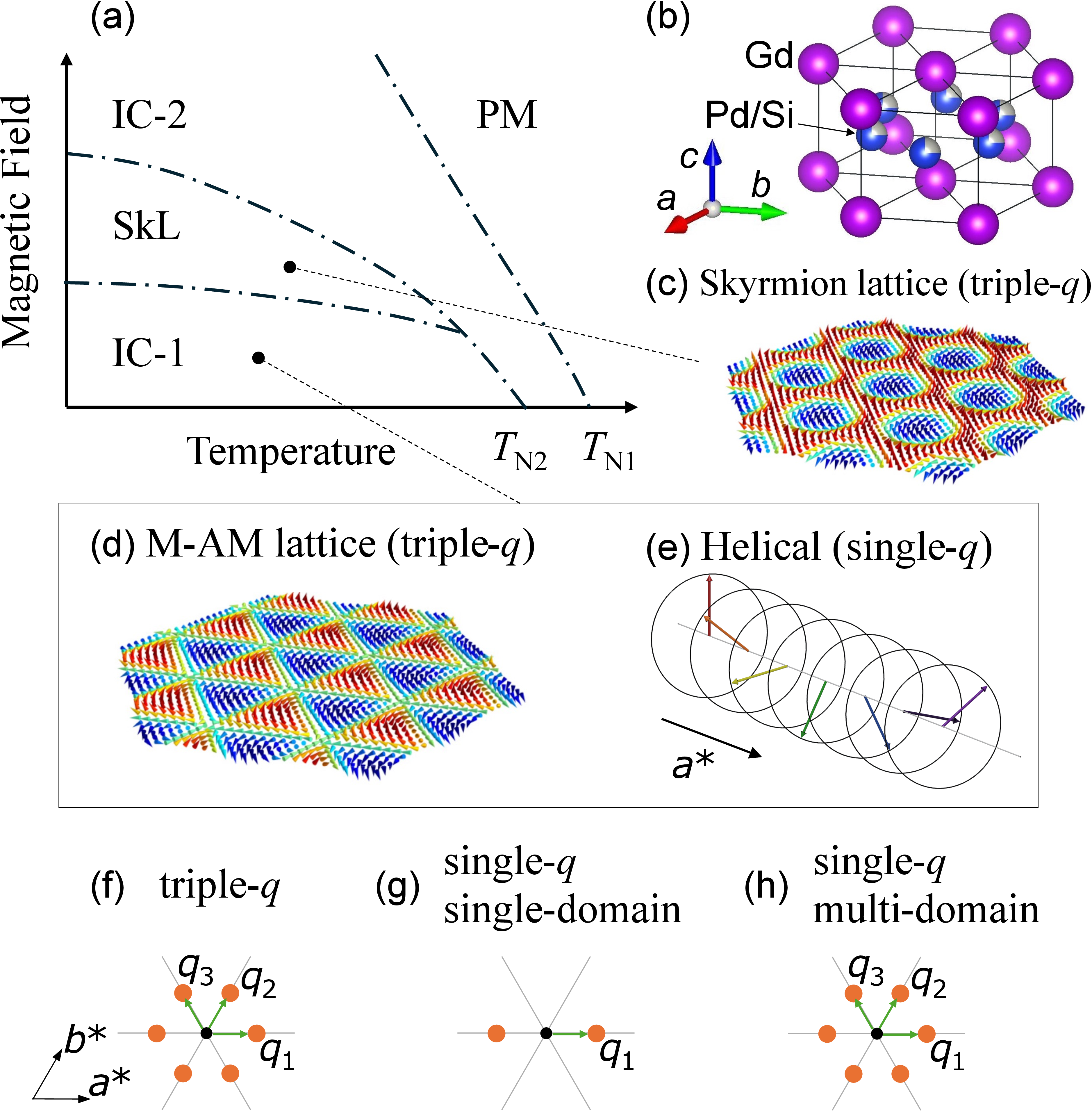}
	\caption{\label{structure}
	(a) Schematic magnetic phase diagram of Gd$_2$PdSi$_3$ for $\mu_0 H \parallel c$.
	(b) Crystal structure of Gd$_2$PdSi$_3$. 
	(c) Illustlation of Skyrmion lattice order. (Reproduced with permission from Ref. \onlinecite{Hirschberger_2020}.)
	(d) Illustration of meron-antimeron (M-AM) order. (Reproduced with permission from Ref. \onlinecite{Hirschberger_2020}.)
	(e) Illustration of a single-$q$ helical order with $q \parallel a^*$.
	Distributions of the Bragg reflections in the neutron experiments of (f) the meron-antimeron (triple-$q$) order, (g) a single-domain helical (single-$q$) order, and (h) a triple-domain helical order, respectively. The black (orange) circles indicate nuclear (magnetic) reflections. The black arrow indicates the reciprocal lattice vector direction. The green arrows indicate the magnetic propagation vectors. The solid lines are guides to the eyes. 
}
\end{figure}

In the present study, we focus on Gd$_2$PdSi$_3$, which has a centrosymmetric hexagonal crystal structure as shown in Fig. \ref{structure}(b). %
The previous study by Kurumaji \textit{et al.}\cite{kurumaji_2019} unveiled the SkL state in the first field-induced phase at low temperatures (Figs. \ref{structure}(a) and (c)). %
In zero field, this system undergoes successive magnetic phase transitions at $T_{\rm{N}1}$ = 22.3 K and $T_{\rm{N}2}$ = 19.7 K \cite{Spachman_2021}, and exhibits two magnetically ordered phases. %
The high- and low-temperature magnetic phases are referred to as IC-2 and IC-1, respectively. %
In both phases, the $q$-vector was determined to be $q=(q,0,0)$, where the incommensurate wavenumber $q$ is approximately 0.14. %
Owing to the six-fold rotational symmetry of the crystal structure, there are also two equivalent $q$-vectors, specifically $(0,q,0)$ and $(q,-q,0)$. %
In the following, we refer to these $q$-vectors of $(q,0,0)$, $(0,q,0)$ and $(q,-q,0)$ as $q_1$, $q_2$ and $q_3$, respectively. %
A previous polarized neutron scattering study revealed that the incommensurate magnetic modulations in IC-2 and IC-1 phases correspond to a transverse sinusoidal and an elliptic screw magnetic modulations, respectively \cite{ju_2023}. %
However, these observations are not enough to conclude whether these magnetic phases have multi-$q$ orders or multi-domain states of single-$q$ orders. %
In the paper reporting the discovery of SkL in this system\cite{kurumaji_2019}, the ground state (IC-1 phase) was considered to be a meron-antimeron (M-AM) lattice structure (Fig. \ref{structure}(d)), which is described by a superposition of three screw-type magnetic modulations, namely a multi-$q$ (triple-$q$) magnetic order. %
A recent muon spectroscopy measurements also reported a meronlike multi-$q$ structure by observing anisotropic spin dynamics in the IC-1 phase \cite{muSR_PRL_2025}. %
By contrast, a study combining neutron scattering and Monte Carlo simulations suggested that the ground state is a single-$q$ elliptic screw magnetic order (Fig. \ref{structure}(e)) \cite{paddison_2022}. %
As illustrated in Fig. \ref{structure}(f) and \ref{structure}(h), the diffraction pattern of the multi-$q$ order is indistinguishable from the multi-domain state of the single-$q$ order if the volume fractions of the magnetic domains are assumed to be equal to each other. %

An effective approach to distinguish between the single-$q$ and multi-$q$ orders is to apply an external perturbation which lifts the degeneracy of the three magnetic modulations. %
In the case of Gd$_2$PdSi$_3$, the three $q$-vectors are interconverted to each other by the three-fold rotational operation about the $c$ axis. %
If the ground state is a multi-domain state of a single-$q$ order with a finite spin-lattice coupling, the application of anisotropic stress perpendicular to the $c$ axis can induce an irreversible change in the volume fractions of the magnetic domains. %
For instance, the application of uniaxial stress, $\sigma$, along the [100] direction can enhance or suppress the $q_2$ domain, where $q_2$ is normal to the [100] direction, while the $q_1$ and $q_3$ domains are suppressed or enhanced accordingly, maintaining equal weights between them.
This can result in a decrease or increase of magnetization along the [100] direction.
In the present study, we employed a compressive uniaxial stress along the [100]([010]) direction and performed magnetization, resistivity and neutron diffraction measurements. %

\section{Experimental details}

We grew single crystals of Gd$_2$PdSi$_3$ with natural Gd and an isotope-enriched $^{160}$Gd$_2$PdSi$_3$ by the floating zone method. %
We used the former crystals for resistivity and magnetization measurements. %
They were cut into rectangular parallelpiped with dimensions of [100] $\times$  [120] $\times$  [001] = 2.64 $\times$ 2.75 $\times$ 1.05 mm$^3$ for resistivity measurement,  2.57 $\times$ 1.06 $\times$ 1.85 mm$^3$ (mass of 33.4 mg) for magnetization measurement. %
For neutron scattering measurements, we used $^{160}$Gd$_2$PdSi$_3$ single crystal. %
We cut a piece which was used for polarized neutron experiment\cite{ju_2023} into a rectangular shape with dimensions of [010] $\times$  [210] $\times$  [001] = 3.94 $\times$ 0.963 $\times$ 2.20 mm$^3$ (mass of 50.2 mg).  

We measured temperature dependence of the magnetization $M$ in an external magnetic field of $\mu_{0}H$ = 0.01 T applied along the [100] direction. %
The measurement was performed using a commercial SQUID magnetometer, Magnetic Property Measurement System (MPMS, Quantum Design inc.), with the uniaxial-stress insert used in Refs. \onlinecite{Nakajima2012,Nakajima2015}, by which we can tune the magnitude of the uniaxial stress even when the sample is at low temperatures.
The direction of the uniaxial stress was parallel to the magnetic field, namely the [100] direction of the sample. %

We also measured resistivity $\rho$ under an application of $\sigma$ along the [100] direction in zero magnetic field. %
The direction of the electric current was set to be [100] direction, and the resistivity was measured by the standard four-probe method. %
Four Au-wires were attached on a $c$-plane by Ag-paste spaced apart from each other.
An AC current with the frequency of 127 Hz was applied, and the resulting voltage was measured by a digital lock-in amplifier (LI5650, NF corp.). %
The sample was mounted in the uniaxial-stress insert used in Ref. \onlinecite{Nakajima2015}, and was loaded into the Physical Property Measurement System (PPMS, Quantum Design inc.).
Similarly to the magnetization measurements, the magnitude of the uniaxial stress was tunable at all temperatures we measured. %

The neutron diffraction measurement was performed at the POlarized Neutron Triple-Axis spectrometer PONTA \cite{5G_PONTA} installed at the 5G beamhole in Japan Research Reactor 3 (JRR-3) of Japan Atomic Energy Agency (JAEA). %
The sample was mounted in a uniaxial-stress insert used in Ref. \onlinecite{Nakajima2012} so as to have the ($H$, 0, $L$) horizontal scattering plane.
The direction of $\sigma$ was parallel to [010], which was normal to the scattering plane (Fig. \ref{neutron}(b)).
The spectrometer was operated in the two-axis mode, and the horizontal collimation was open-80'-80'.
An incident neutron beam with the wavelength of 2.36 \AA  was obtained by a PG (002) monochromator. %
A part of the measurement was carried out using a scintillator-type two-dimensional position sensitive detector (2D-PSD) which has the neutron-sensitive area of 256 mm $\times$ 256 mm  \cite{nakamura_2012}. %
A 2D-PSD was placed at $2\theta$ = 39.3 deg. with its normal vector pointing from the center of 2D-PSD to the sample position, which was confirmed by the $2\theta$ angle of 100 nuclear Bragg reflection. %
The sample-detector distance was set to 764 mm. %
As a result, the vertical and horizontal acceptance of the 2D-PSD was $\pm$9.5 deg. %
The detector efficiency was calibrated by measurements on a standard vanadium sample. %

\section{Results and discussions}
\subsection{Magnetization and resistivity measurements under $\sigma$}

\begin{figure}[t]
	\includegraphics[keepaspectratio, width=7.5cm]
	{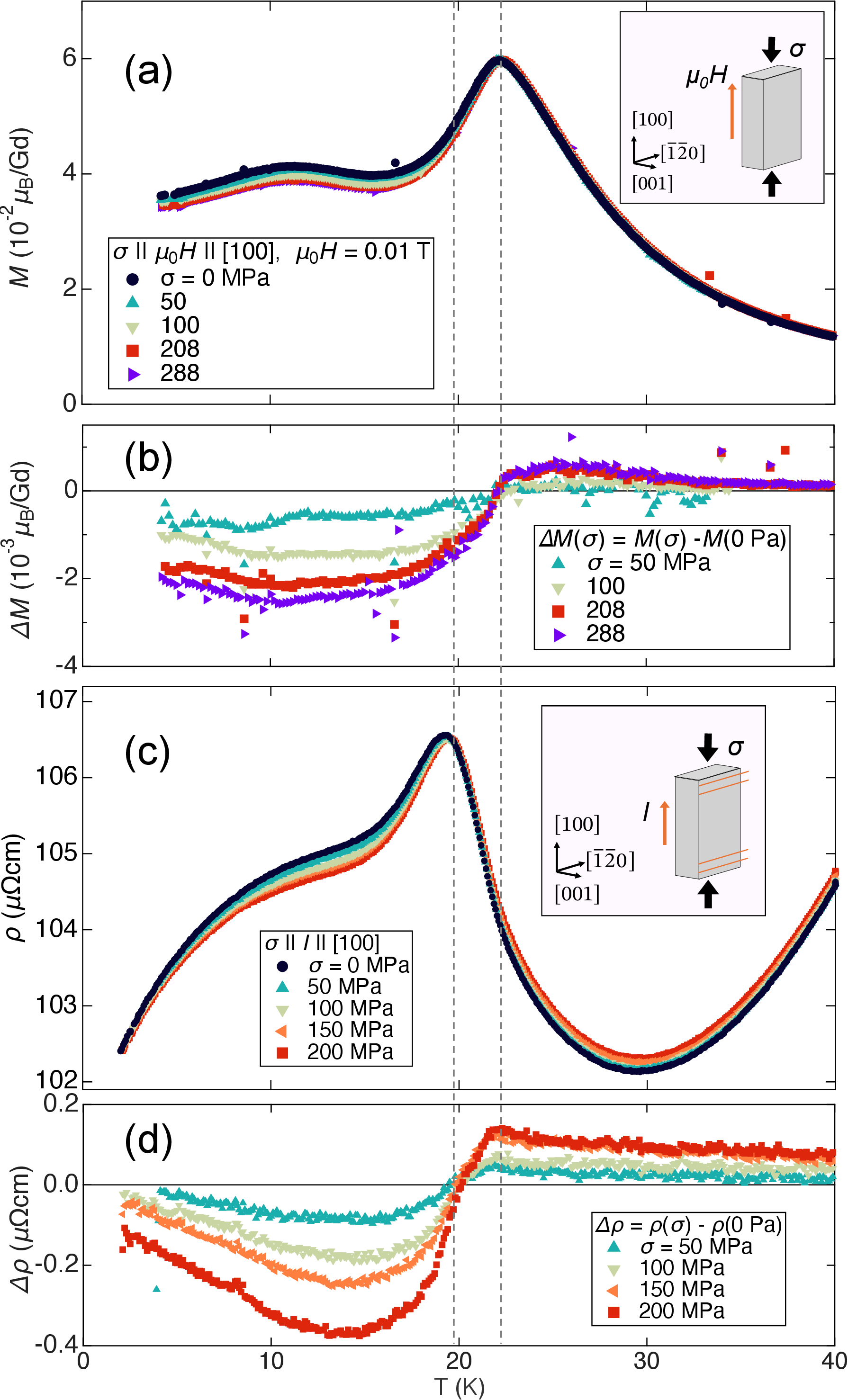}
	\caption{\label{tdep} 	
	(a) Temperature dependence of magnetization $M$ at $\mu_0H$ = 0.01 T and $\sigma$ = 0, 50, 100, 208, and 288 MPa. $M \parallel H \parallel \sigma\ \parallel [100]$. 	
	(b) Temperature dependence of the difference of magnetization $\Delta M$. 	
	(c) Temperature dependence of resistivity $\rho$ at $\mu_0H$ = 0 T and $\sigma$ = 0, 50, 100, 150, and 200 MPa. Electric current $I$ is applied along $\sigma  \parallel [100]$.	
	(d) Temperature dependence of the difference of resistivity $\Delta \rho$. The measurements were performed under cooling condition.  The dashed lines indicate the literature value of $T_{\rm{N}1}$ = 22.3 K and $T_{\rm{N}2}$ = 19.7 K, respectively.\cite{Spachman_2021} A schematic illustration of the geometry for each measurements are shown in inset.}
\end{figure}

\begin{figure}[t]
	\includegraphics[keepaspectratio, width=7.5cm]
	{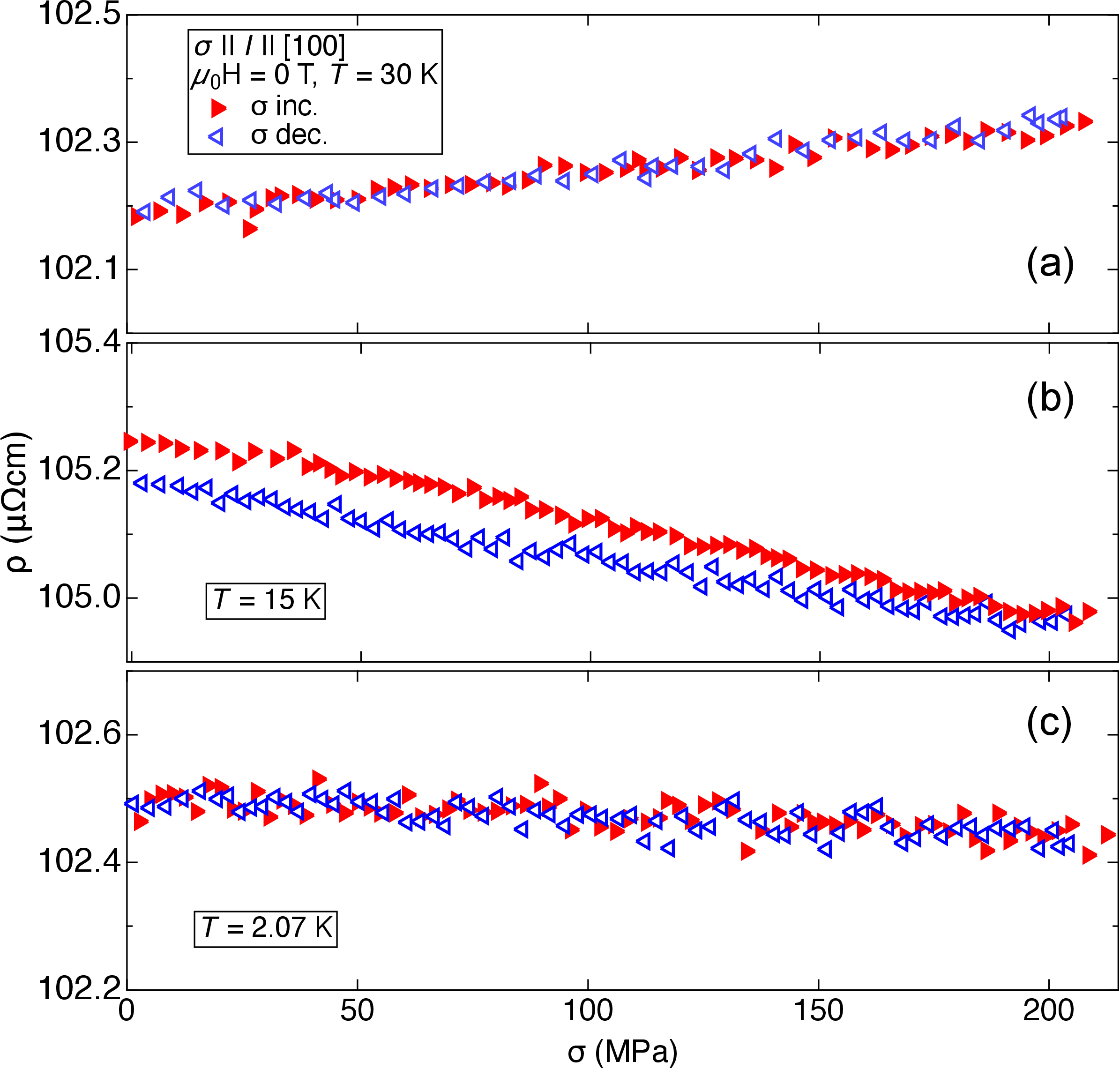}
	\caption{\label{sigdep} Uniaxial stress dependence of $\rho$ at $\mu_0H$ = 0 T and $T$ = (a) 30 K, (b) 15 K, and (c) 2.07 K, respectively. $I \parallel \sigma \parallel [100]$. The closed (open) symbol indicates the data measured under $\sigma$ increase (decrease).}
\end{figure}

Figure \ref{tdep}(a) shows the temperature dependence of the magnetization $M$ in a magnetic field of 0.01 T applied parallel to the [100] direction. %
At ambient pressure, the magnetization is suppressed and start decreasing below at around 22 K, which corresponds to the phase transition from paramagnetic (PM) phase to the IC-2 phase, $T_{\rm{N}1}$. %
Gd$_2$PdSi$_3$ undergoes the magnetic transition to IC-1 at $T_{\rm{N}2}$. %
According to the previous studies, the phase transition at $T_{\rm{N}2}$ results in a local maximum in the magnetization along the $c$ axis \cite{kurumaji_2019,ju_2023}, while no clear anomaly is observed in the $a$-axis magnetization. %
The present results are well consistent with the previous result. %

We then applied uniaxial stress $\sigma$ = 50, 100, 208, and 288 MPa along [100] direction, revealing that  below $T_{\rm{N}1}$, $M$ is slightly suppressed as the applied stress is increased. %
This can be seen in the temperature dependence of $\Delta M$, which is a difference between $M$ under $\sigma$  and $M$ at ambient pressure, shown in Fig. \ref{tdep}(b). %
Since the magnetization is assumed to be isotropic in the $ab$ plane at ambient pressure, $\Delta M$ can be interpreted as a $\sigma$-induced in-plane anisotropy of $M$. %
In the PM phase, $\Delta M$ shows no significant $\sigma$ dependence. %
On the other hand, $\Delta M$ shows a kink anomaly at $T_{\rm{N}1}$ and turns to be negative in the ordered phases. %
A remarkable feature in the temperature dependence is that $\Delta M$ shows a local minimum between 10 and 15 K and then the magnitude of $\Delta M$ becomes smaller with decreasing temperature. %

A similar tendency was also observed in the resistivity measurements under $\sigma$ in zero magnetic field.
The uniaxial stress was applied along [100] direction in the same manner as the magnetization measurements. %
Figure \ref{tdep}(c) shows the temperature dependence of the resistivity $\rho$.
At ambient pressure, $\rho$ shows a marked increase below 30 K. 
While $\rho$ does not show a clear anomaly at $T_{\rm{N}1}$, it starts to decrease below $T_{\rm{N}2}$. %
By applying the uniaxial stress, $\rho$ is slightly increased in PM phase, while it is suppressed in the ordered phases. %
Similarly to the magnetization measurements, the temperature dependence of $\Delta \rho$, which is a difference between $\rho$ under $\sigma$ and $\rho$ at ambient pressure, is calculated as shown in figure \ref{tdep}(d). %
While the magnetic phase transition at $T_{\rm{N1}}$ was not clearly seen in the temperature dependence of $\rho$, we observed a kink anomaly of $\Delta \rho$ at around $T_{\rm{N1}}$. %
The slight difference between the temperatures at which the kink anomalies were observed in $M$ and $\rho$ might be caused by the experimental setup, which could have temperature gradient between the sample and the thermometer.
$\Delta \rho$ also shows a local minimum at around 15 K and then $\Delta \rho$ approaches zero as the temperature is decreased. %

We note here that the stress-induced anisotropy in resistivity was also reported in other magnetic systems. %
For instance, iron-based superconductors and their parent compounds are known to have multiple magnetic domains\cite{Chu_science_2010}. %
Most of them have tetragonal crystal structures, and the four-fold rotational symmetry was broken by the magnetic orders. %
By applying a weak uniaxial stress perpendicular to the $c$ axis, the volume fractions of the magnetic domains were easily controlled owing to the spin-lattice coupling in the single-$q$ magnetic domains. %
This results in the resistivity anisotropy that is, in most cases, maintained across the whole temperature range of the magnetic phase \cite{Chu_science_2010,Ba122_detwin}. %
Another example is the commensurate antiferromagnet CeRh$_2$Si$_2$, which exhibits both single-$q$ and multi-$q$ phases in zero field \cite{CeRh2Si2_PRB_2023}. %
In this system, the stress-induced resistivity anisotropy was enhanced as temperature decreased in the whole temperature range of the single-$q$ magnetic phase. %
As compared to these previous results, the present results imply that the $\sigma$-induced anisotropy in resistivity in Gd$_2$PdSi$_3$ is not attributed to a simple repopulation of the single-$q$ magnetic domains. %

\begin{figure*}[tbp]
\centering
	\includegraphics[keepaspectratio, scale=.22, width=0.7\linewidth]
	{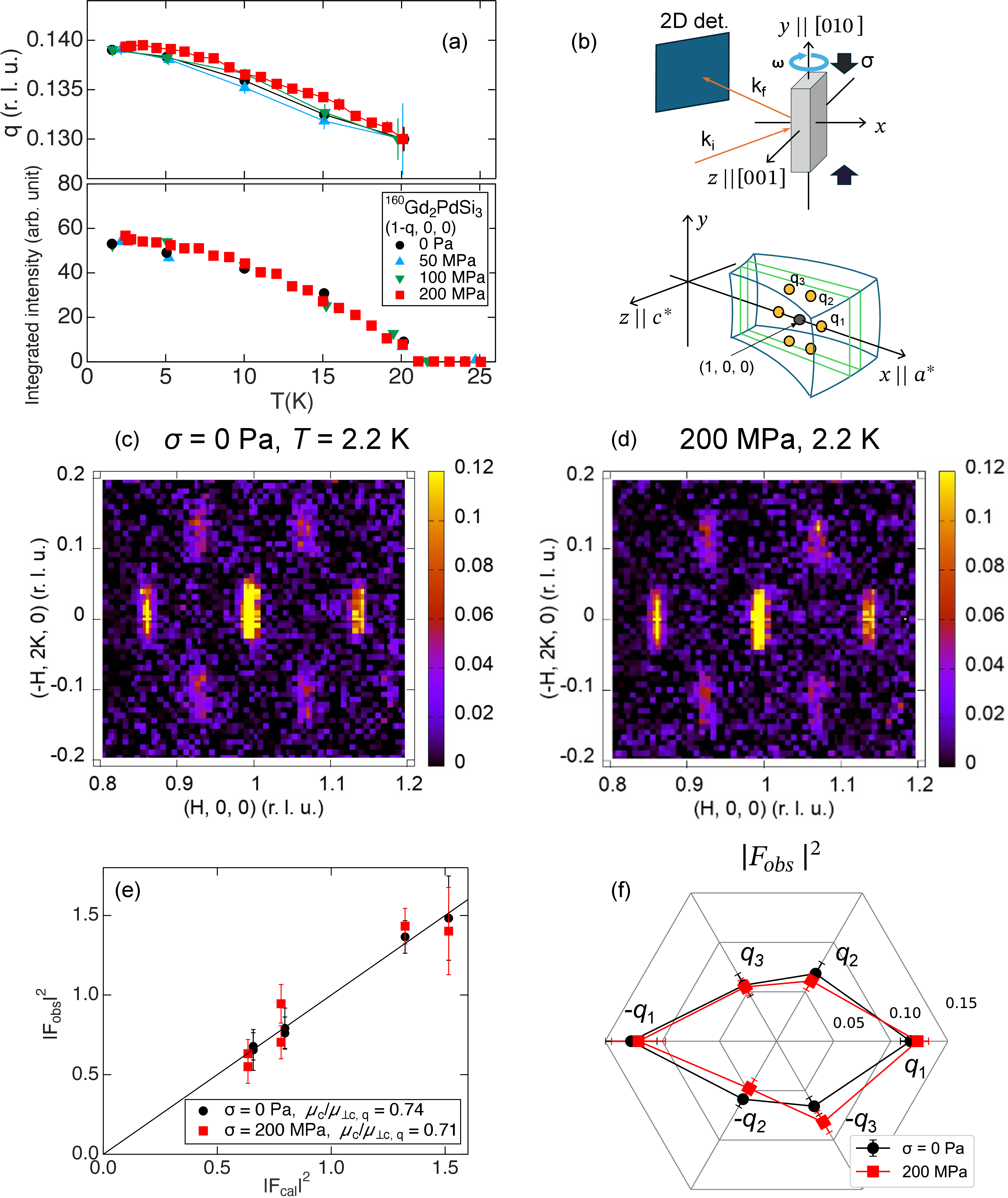}
	\caption{\label{neutron}
	(a) Temperature dependence of the magnetic propagation vector $q$  and the integrated intensity of (1-$q$, 0, 0) reflection measured at $\sigma$ = 0, 50, 100, and 200 MPa.
	(b) Schematic illustrations of the experimental geometry in (top) real and (bottom) reciprocal space. A box at the origin represents the sample. The black arrow indicates the direction of $\sigma$. The orange arrows indicate incident and scattered neutron, $k_i$ and $k_f$. The light-blue arrow indicates an angle of sample rotation, $\omega$. The blue square indicates 2D-PSD. The yellow (black) spheres represent magnetic (nuclear) Bragg reflections. A volume surrounded by blue solid lines is observed trajectory. A portion of volume indicated by light-green lines is integrated along $c^*$ to obtain (c) and (d).
	(c),(d) Intensity maps in ($H$, $K$, 0) plane at around (1, 0, 0) measured at $T$ = 2.2 K and $\sigma$ = 0 and 200 MPa, respectively. The color scale indicates the intensity.
	(e) $|F_{cal}|-|F_{obs}|$ plot for the magnetic reflections observed in (c) and (d). The solid lines are the result of linear fitting. $\mu_c/\mu_{\perp c, q}$ is the ellipticity of the screw. The integrated intensities of the magnetic Bragg reflections around (1, 0, 0) were obtained by integrating the relevant regions in (c) and (d).
	(f) Spider chart for $|F_{obs}|^2$ of the six magnetic reflections around 100 reflection derived from (c) and (d).}
\end{figure*}

This conclusion is further corroborated by the isothermal $\sigma$ dependence.
Generally, the orientation of the single-$q$ magnetic domains and concomitant changes in the bulk properties are irreversible.
In fact, significant hysteresises in the $\sigma$ dependence of $M$ and $\rho$ were observed only in the single-$q$ phase of CeRh$_2$Si$_2$, where their values were maintained upon decreasing $\sigma$ \cite{CeRh2Si2_PRB_2023}. 
These irreversible responses are direct evidence of the $\sigma$-induced repopulation of the single-$q$ magnetic domains. %
We thus investigated the $\sigma$ dependence of $\rho$ in Gd$_2$PdSi$_3$ in isothermal conditions to search for similar hysteretic behavior. %
 As shown in Figure \ref{sigdep}(a), at $T$ = 30 K in the PM phase, $\rho$ linearly increases as $\sigma$ increases.
On the other hand, in the IC-1 phase, $\rho$ linearly decreases as $\sigma$ increases as shown in Figure \ref{sigdep}(b) and (c).
The slope is close to zero at $T$ = 2.07 K. %
These $\sigma$-induced changes of $\rho$ agree with those observed in the temperature dependence measurements. %
Therefore, the $\sigma$ dependence of $\rho$ also indicates that the IC-1 phase is not a single-$q$ order.
We note that a relatively steep slope and a small hysteretic feature at 15 K may indicate that the electronic state of the IC-1 phase (and then the magnetic order itself) can be deformed by the in-plane uniaxial stress at moderate temperature.

\subsection{Neutron diffraction measurements}

To directly observe the multi-$q$ nature of the IC-1 phase, we performed single-crystal neutron diffraction measurements on an enriched $^{160}$Gd$_2$PdSi$_3$ at ambient pressure and under the $\sigma$ of up to 200 MPa applied along [010].
We focused on the magnetic Bragg reflections at (1,0,0) $\pm$ $q_1$, $q_2$, or $q_3$.
First, we measured the temperature dependence of the integrated intensity and the propagation vector $q$ of the magnetic Bragg reflection at $(1-q, 0, 0)$ under the application of $\sigma$ = 0, 50, 100, and 200 MPa as shown in Fig. \ref{neutron}(a). %
For these measurements, the spectrometer was operated in the two-axis diffraction mode with a standard single-channel detector. %
 At ambient pressure, the integrated intensity at $(1-q, 0, 0)$ appears at around $T_{\rm{N}1}$, increases as temperature decreases, and saturates at the lowest temperature. %
The $q$ value gradually increases from approximately 0.13 to 0.14 as temperature decreases. %
By applying uniaxial stress up to 200 MPa, we observed no significant change in both $q$ and the integrated intensity. %
If the IC-1 phase is a single-$q$ phase with a finite spin-lattice coupling, an enhancement or reduction of the integrated intensity at $(1-q, 0, 0)$ would be expected under applied $\sigma$. %
The observed data would be interpreted that the IC-1 phase is a multi-$q$ phase, which is robust against the application of the uniaxial stress. %
However, it might be possible to assume that the system was already in the detwinned single-$q$ state characterized by $q_1$ by some residual stress applied when preparing the sample. %

To exclude this possibility, we measured not only the magnetic reflections corresponding to $q_1$, but also those to $q_2$ and $q_3$, which were located out of the horizontal scattering plane by using a 2D-PSD, as shown in Figure 4(b). %
We measured three dimensional distributions of the intensities in the reciprocal space by rotating the $\omega$ angle of the sample, and extracted the intensity maps for the $(H,K,0)$ plane. %
Figure 4(c) shows the intensity map measured at ambient pressure at 2.2 K. %
We observed six magnetic Bragg reflections around (1, 0, 0). %
The integrated intensity of each reflections can be well explained by the model composed of the three screw-type magnetic modulations with equal amplitudes, as shown by black circle in $|F_{cal}|-|F_{obs}|$ plot (Fig. 4(e)).
The ellipticity of the screw $\mu_c/\mu_{\perp c,q}$ is estimated to be 0.74 from the fit, which is consistent with the previous polarized neutron scattering measurements\cite{ju_2023}. %
Figure 4(d) shows the intensity map measured at $\sigma$ = 200 MPa at 2.2 K.
All the satellite reflections around (1, 0, 0) are observed even under the application of the uniaxial stress. %
The integrated intensity of each reflection under the uniaxial stress and at ambient pressure are nearly the same within the error bars as shown in Fig. 4(f). %

From these results, we suggest that that the IC-1 phase is a multi-$q$ order, and that the single-$q$ order is not likely. %
Taking into account the fact that the intensities of the six satellite reflections near the 100 reflection were explained by the model having three screw modulations with the equal amplitudes, it would be reasonable to assume a triple-$q$ meron-antimeron lattice for the IC-1 phase. %
We would note that the above consideration is based on an assumption that all the magnetic modulations ($q_1$, $q_2$, and $q_3$) have the same screw-type symmetry, which is suggested by the previous polarized neutron scattering experiment\cite{ju_2023}. 
However, there are several multi-q orders in which multiple magnetic modulations having different symmetries (screw, cycloid, or sinusoidal) combined\cite{Yoshimochi_2026,Ishiwata_2020}.
If this is also the case for the IC-1 phase, it might be possible to consider a two-domain state of an anisotropic double-$q$ structure having a robustness in magnetic intensities to $\sigma$ applied parallel to $a$ axis. %
To address this point, other experimental techniques having spatial resolution, such as resonant x-ray diffraction with focused beam, will be necessary in the future. %
We also note here that the ellipticity of the screw $\mu_c/\mu_{\perp c,q}$ is slightly reduced to 0.71 by application of the uniaxial stress. %
This implies that the application of $\sigma$ slightly affected the magnetic anisotropy while keeping the multi-$q$ structure. %

\section{Conclusion}
To summarize, we performed magnetization, resistivity, and neutron elastic scattering measurements under uniaxial stress on single crystals of Gd$_2$PdSi$_3$. 
In the IC-1 phase, we observed a suppression of magnetization and resistivity under uniaxial stress applied along the $a$ axis ($\sigma \parallel a$). 
However, considering their temperature and stress dependencies, this behavior is difficult to explain by the domain repopulation of a single-$q$ order. 
Furthermore, the neutron diffraction measurements with 2D-PSD unequivocally show the existence of the magnetic reflections corresponding to the three $q$-vectors even under application of the uniaxial stress of 200 MPa.
This excludes the possibility of the single-$q$ order, suggesting the IC-1 phase is a multi-$q$. 
We would note that this observation is in contrast to EuAl$_4$, another centrosymmetric skyrmion host, in which a single-$q$ single domain state could be achieved under a uniaxial stress of approximately 50 MPa, and even the transition temperature could be tuned\cite{Gen_2026}.

Regarding the microscopic model for this system, the dominant interaction in Gd$_2$PdSi$_3$ can be estimated by comparing these findings with the calculated magnetic phase diagram. 
Among the proposed theories \cite{PRL_Okubo_MonteCarlo,NCom_Leonov_frustratedmag,PRB_Hayami_biquadratic,PRL_Wang_2DEG,Sk_AnisoExchange_NJP_2021,Sk_dipolar_PRB_2021}, only the one considering a biquadratic term of the exchange interaction predicts a multi-$q$ magnetic order as the ground state \cite{PRB_Hayami_biquadratic}. 
This term is an effective higher-order term of the exchange interaction of localized spins mediated by itinerant electrons, derived from the perturbative expansion with respect to the exchange coupling in the Kondo lattice model. %
Our results thus suggest that such higher-order interaction plays an important role in the formation of the magnetic ordered phases in Gd$_2$PdSi$_3$.
Additionally, temperature dependence of $|\Delta M|$ and $|\Delta \rho |$ showed local maxima at an intermediate temperature of approximately 15 K. 
This suggests that the magnetic structure may undergo deformation in this intermediate temperature region. 
To clarify the nature of the pronounced $\sigma$ dependence, more detailed neutron diffraction measurements of all six magnetic satellite reflections at various temperatures are required.

\begin{acknowledgments}
The neutron scattering experiments at PONTA in JRR-3 were carried out along the proposals (No. 22401 and 23401, respectively) and partly supported by the Institute for Solid State Physics of the University of Tokyo. %
A part of the magnetization measurements were performed by using MPMS at the user laboratory of the Comprehensive Research Organization for Science and Society.
This work was supported by JSPS KAKENHI (Grants No. 25K17338). %
The images of the crystal and magnetic structures in this paper were depicted using the software VESTA \cite{VESTA} developed by K. Momma. %

\end{acknowledgments}

\bibliography{Gd2PdSi3_uniax_aps}

\end{document}